\documentclass[aip,reprint]{revtex4-1}

\usepackage{amsmath,amssymb,graphicx}
\usepackage{graphicx}% Include figure files

\draft % marks overfull lines with a black rule on the right

\begin{document}

\title{Veselago lens by photonic hyper-crystals }

\author{Zun Huang}
\email{ zun@purdue.edu}
\author{Evgenii E. Narimanov}

\affiliation{School of Electrical and Computer Engineering, Purdue University, West Lafayette, Indiana 47907, USA}
\affiliation{Birck Nanotechnology Center, Purdue University, West Lafayette, Indiana 47907, USA}

\date{\today}

\begin{abstract}
Based on the recent concept of the photonic hyper-crystal -- an artificial optical medium that combines the properties of hyperbolic materials and photonic crystals, we present the imaging system functioning as a Veselago lens. This planar lens shows a nearly constant negative refractive index and substantially reduced image aberrations, and can find potential applications in photolithography and hot-spots detection of silicon-based integrated circuits. 

\end{abstract}

\maketitle %\maketitle must follow title, authors, abstract and \pacs

%%%%%%%%%%%%%%%%%------------------========== Body of paper =======------------------%%%%%%%%%%%%%%

%%******************************************************  Sec. I  ****************************************************** %%

%\section{Introduction}

Veselago lens, originally introduced in 1968 \cite{Veselago1968}, is a planar slab capable of focusing divergent homocentric electromagnetic beams based on the  phenomenon of negative refraction, with the potential for subwavelength resolution imaging \cite{Pendry2000, Blaikie2005, ZhangScience05}. The planar design of Veselago lens is also advantageous for nano-fabrication and has shown innovative applications in improving the performance of absorbance-modulation optical lithography \cite{Blaikie2011}. 

Thanks to the rapid development of metamaterials \cite{DRSmithScience04}, artificially nanostructured composites embedded with periodic subwavelength inhomogeneities, Veselago lens can be realized by negative index metamaterials (NIMs) in a wide range of electromagnetic spectrum \cite{SchultzScience2001, Chuang, ZhangScience2004, DRSmithScience04, ShalaevNature06, ZhangScience08}. However, a double resonance overlapping window is required to obtain simultaneously negative permittivity and permeability, leading to inevitable resonance losses and technical difficulties in design and fabrication.

The core foundation of Veselago lensing relies on the extraordinary effect of negative refraction, when the direction of refracted energy flow (represented by Poynting vector $\vec{\bold S}$) stays at the same side of surface normal in different media. In the quest of negative refraction, a variety of significant alternatives beyond NIMs have been proposed and experimentally verified, such as photonic crystals (PCs) with folded band diagrams \cite{Notomi, PendryAANR, SridharNature, SoukoulisNature, SoukoulisPRL, Anand2004}, crystal structures of plasmonic waveguides \cite{Shvets2003, Fan2006, Lezec2007, Polman2010, XuNature2013}, hyperbolic metamaterials (HMMs) \cite{SmithPRL2003, SmithAPL2004, HoffmanNature07, SoukoulisPRB09} and etc. 

However, a Veselago lens accomplished by either PCs or HMMs fails to yield an abberation-free image. There, the mutual angle between $\vec{\bold S}$ and $\vec{\bold k}$ are dependent of the band diagram anisotropy \cite{PendryAANR} or the propagating $\vec{\bold k}$ direction with respect to major optical axis \cite{SmithAPL2004}. Consequently, the index of refraction is strongly angle-dependant causing an imperfect focusing. For example, waves emanating from a point source with large tangential momenta become defocused through a Veselago lens made of HMMs \cite{SmithAPL2004}. 
 
In this Letter, we report a new approach to achieve Veselago lens based on the recently introduced concept of photonic ``hyper-crystals" (PHCs) \cite{NarimanovArXiv}, where Bragg scattering and band gaps persist even in the metamaterial limit. Our approach demonstrates substantially improved image quality and relatively low losses. As an example of application, we show the PHC based Veselago lens can be used to diagnose hot-spots in silicon-based integrated circuits.

%%******************************************************  Sec. II  ****************************************************** %%

%\section{Theory, Methodology and Results }

The relationship between the unit cell size $D$ and free-space wavelength $\lambda_0$ in an artificial optical media defines the electromagnetic properties and the corresponding physics and phenomena. PCs \cite{Wenshan} and optical metamaterials \cite{Joannopoulos} represent two limits in these parameter spaces. In PC regime where $D \gtrsim \lambda_0$, the light propagation is featured by energy bandgaps due to Bragg scattering, which is also responsible for the unlocalized optical Tamm states \cite{Joannopoulos}. On the other hand, in metamaterial regimes where $D \ll \lambda_0$, the macroscopic electromagnetic response can be described in terms of averaged permittivity ($\epsilon$) and permeability ($\mu$) tensors \cite{Wenshan}. By manipulation of the unit cell components and geometry, a broader degree of controlling the optical response can be achieved, such as negative refraction and negative refractive index. For most optical materials in these two regimes, the wave number $k$ of propagating modes is usually limited within the same order of magnitude as the corresponding free-space values, i.e. $k D \ll 1$. As a result, subwavelength features of an object corresponding to large wave numbers cannot be retrieved by a conventional Fourier imaging device since those information carried by evanescent waves are already vanished before reaching a detector. 

%
%---------------------- Fig. 1 ---------------------
\begin{figure}
\centering
\includegraphics[width=8.5cm]{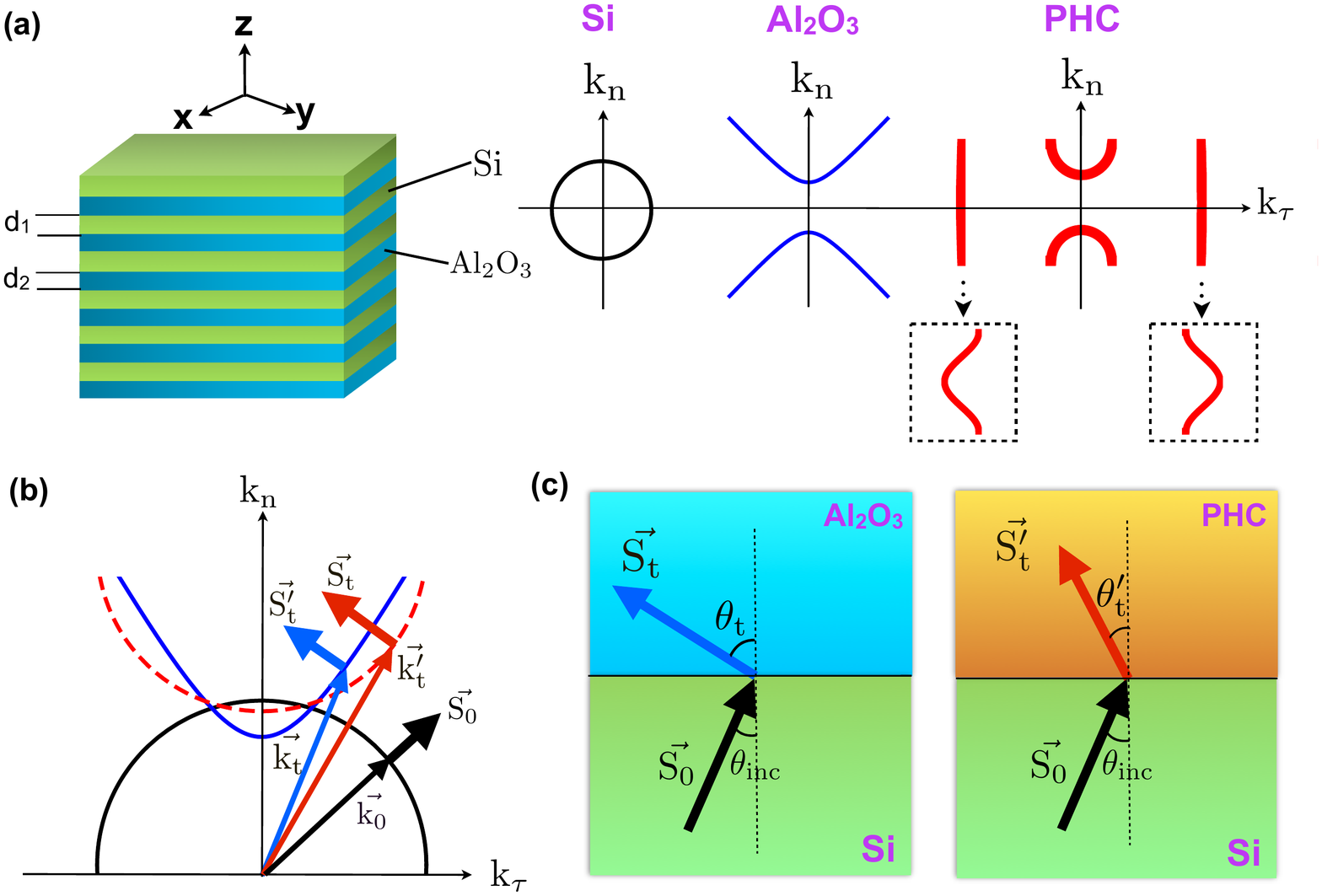} 
%\noindent
\caption{\noindent Schematics of the proposed Veselago lens by a planar PHC composed of interleaved Si-$\rm Al_2O_3$  layers and the associated IFCs. (a) The planar layered structure of PHC and dispersion functions of Si (as dielectric, black circle), sapphire (as hyperbolic medium, blue hyperbola) and PHC (red curves). The single period of the PHC satisfies the metamaterial limit: $ d_1 +  d_2 \ll \lambda_0 $. The insets (dashed boxes) show two enlarged narrow propagating bands corresponding to the nearly vertical lines in the dispersion function of PHC. (b) The propagating directions of phase ($\vec {\bold k}$) and energy ($\vec {\bold S}$, normal to the curves) of a TM polarized plane wave in three different media shown by the corresponding IFCs. The bold and thin arrows represent Poynting vectors ($\vec {\bold S}$) and wave vectors ($\vec {\bold k}$) in Si ($\vec {\bold S_{\rm0}}$, $\vec {\bold k_{\rm 0}}$), sapphire ($\vec {\bold S_{\rm t}}$, $\vec {\bold k_{\rm t}}$) and PHC ($\vec {\bold S_{\rm t}'}$, $ \vec {\bold k_{\rm t}'}$) respectively. Note that the red dashed curve has the same magnitudes of the curvatures as the black circle. (c) The negative refraction of a TM polarized plane wave at the interface of Si (with incident angle $\theta_{\rm inc}$) and two other media: sapphire (left, with refracted angle $\theta_{\rm t}$) and PHC (right, with refracted angle $\theta_{\rm t}'$). While in sapphire $\theta_{\rm t} $ is generally unequal to $\theta_{\rm inc}$, PHC preserves the refracted angle $\theta_{\rm t}' = \theta_{\rm inc}$. The dotted line shows direction of the normal to the surface.}
\label{fig1}
\end{figure}
%-----------------------------------------------------
%

There is however a special class of metamaterials - HMMs, which can support extremely large wave numbers in the propagating modes, leading to a broadband super-singularity in photonic density of states \cite{ENPRL2010} and a wide range of novel applications - from super-resolution imaging by hyperlens \cite{ENOE2006, ZhangScience2007} to enhanced spontaneous emission \cite{ENAPL2012, ENOL2010}. When the eigenvalues of the permittivity tensor in two orthogonal directions ($\epsilon_\tau = \epsilon_x=\epsilon_y$ and $\epsilon_n = \epsilon_z$) take opposite signs, the dispersion function of constant frequency $\omega$ becomes hyperbolic for a TM-polarized wave: $k_n^2- (- \frac{\epsilon_\tau}{\epsilon_n})k_\tau^2 =  \epsilon_\tau \frac{\omega^2}{c^2} $. Aside from the metamaterial realizations of hyperbolic media, many natural materials also show the hyperbolic response. For example, in far-IR range relevant for  spectroscopy and thermal imaging, sapphire behaves as a natural hyperbolic medium \cite{sapphire}, as shown by its iso-frequency curve (IFC) in Fig. 1 (a). 

PHC is essentially a photonic crystal with a deep subwavelength unit cell that includes elements exhibiting hyperbolic dispersion \cite{NarimanovArXiv}, and  combines the properties of both HMMs and PCs.  First, the wave dispersion in PHC is much more complicated than either of its cell elements, which exhibits multiple allowed and forbidden bands in both frequency and momentum domains - as shown in Fig. 1(a). Furthermore, Bragg scattering of the high-$k$ waves in bandgaps can excite strong localized subwavelength surface modes with substantially low loss. Those new features could exert significant impact on the electromagnetic properties of PHC and imply intriguing applications. In particular, the use of PHC can resolve the strong image aberration problem that plagues the HMM-based Veselago lens. 

Fig. \ref{fig1} shows a PHC composed of Si-$\rm Al_2O_3$ alternating layered structure with a subwavelength period $d_1+d_2 \ll \lambda_0$, where $d_1$ and $d_2$ are the layer thicknesses of Si and sapphire ($\rm Al_2O_3$) respectively. With appropriate selection of $d_1$ and $d_2$, the first and also the widest propagating band in PHC can be modulated to resemble a half-circle. While negative refraction can be observed in both PHC and a metallic-type ($\epsilon_\tau > 0$ and $\epsilon_n < 0$) hyperbolic medium \cite{SmithPRL2003, SmithAPL2004, HoffmanNature07, SoukoulisPRB09}, the latter is unable to focus a bunch of homocentric incident beams. This is because the refracted angles in such hyperbolic medium are dependent on the incident angles. However, this behavior is completely different in PHC. As shown in Fig 1(c), the refracted angles in PHC are always equal to the incident angles, indicating a perfect focusing of all refracted beams and a constant refractive index independent of incident angles. As a result, a planar hyperbolic medium used to image a point source results in an imperfect focusing  (Fig. 2(a)), but a planar PHC is capable of focusing waves emanating from a point source with aberration-free images (Fig. 2(b)). Even in the case of lossy sapphire,  a nearly perfect focusing is still attainable. 

%
%----------------------  Fig. 2 ---------------------
\begin{figure}
\centering
\includegraphics[width=8.5cm]{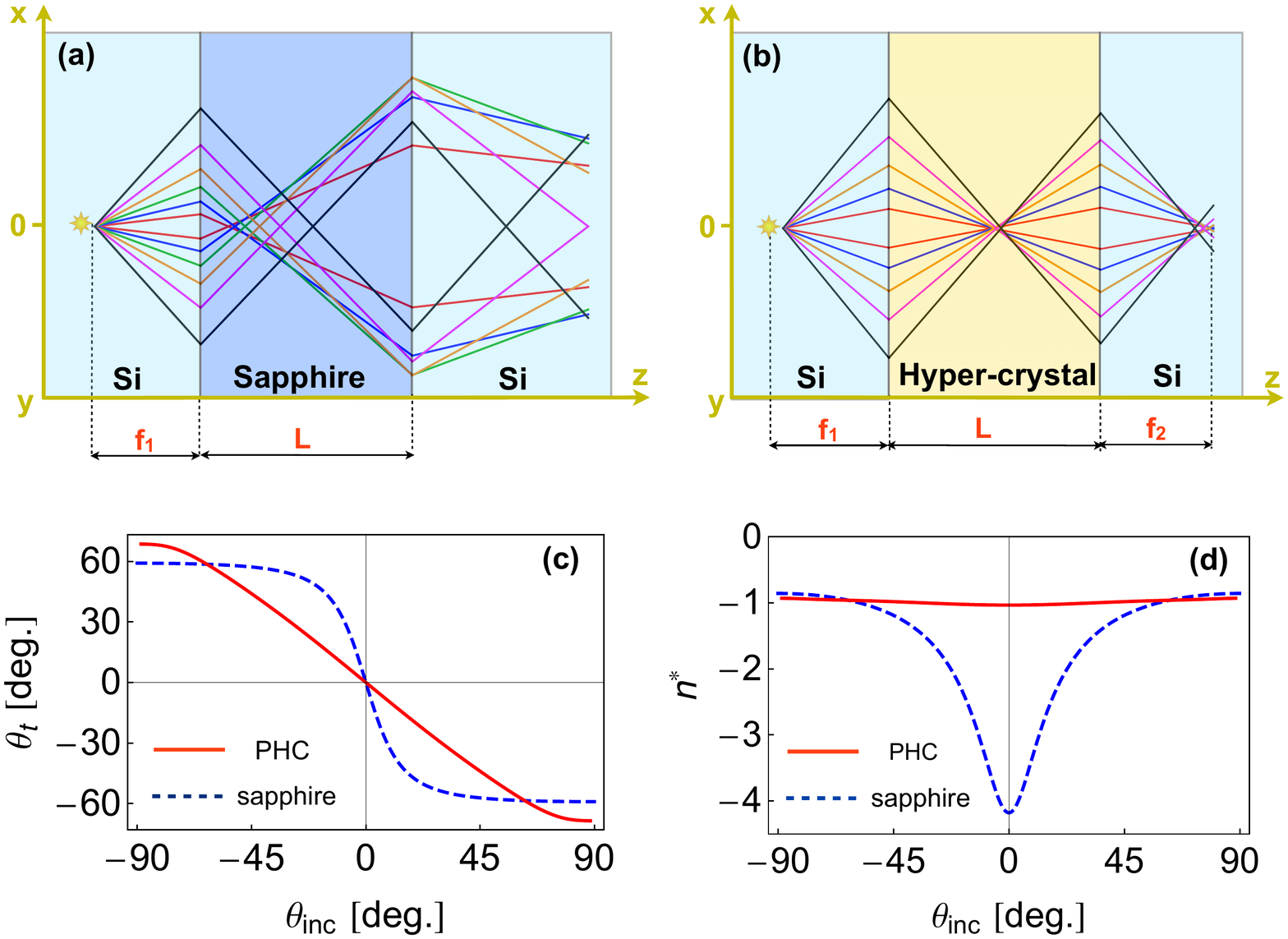} 
%\noindent
\caption{\noindent (a-b) Ray trajectories from a point slot source through two slab media: sapphire (a) as hyperbolic medium, and PHC (b) as the proposed Veselago lens. In both cases the distance of source from the slab is $\rm f_1$ = 51.45 $\mu \rm m$ and the length of slab medium is L = $2 \rm f_1$. In panel (b), the incident beams are focused at the exit plane with a distance $\rm f_2 \approx \rm f_1$ from the lens,  while in panel (a) no focusing is observed. (c) The refraction angle $ \theta_{\rm t}$ as a function of incident angle $ \theta_{\rm inc} $ when a TM polarized plane wave enters the sapphire (blue dashed) and PHC (red solid) respectively from Si ambient at $\lambda_0$ = 20 $\mu \rm m$; (d) The relative index $n^*$ defined by Snell's law as a function of $\theta_{\rm inc}$ for sapphire (blue dashed) and PHC (red solid). 
}
\label{fig2}
\end{figure}
%------------------------------------------------------
%

For optimal imaging performance, the component layer thicknesses ($d_1$ and $d_2$) in PHC are obtained from the following boundary conditions in the lossless limit: (1) the IFC of PHC hits the bandgap edge when the tangential momentum $k_\tau = \sqrt{\epsilon_{\rm 0}} \omega/c$, where $\epsilon_0$ is the ambient dielectric permittivity; (2) the magnitude of the IFC curvature of PHC is equal to that of the ambient at $k_\tau = 0$. These conditions are shown by the following equations:
\begin{eqnarray}
\rm k_n|_{\! \; \rm k_\tau=\sqrt{\epsilon_0}\frac{\omega}{c}} \; &= \;{\pi \over \rm d_1+\rm d_2}, \\
\frac{\mathrm d^2 \rm k_n}{\mathrm d \rm k_\tau^2}|_{\! \; \rm k_\tau=0} \; &= \; \frac{1}{\sqrt{\epsilon_0}\frac{\omega}{c}}.
\label{optimal}
\end{eqnarray}

Considering the planar PHC discussed above to image a point slot source in Si ambient, the refracted angle $\theta_{\rm t}$ can be calculated as a function of  the incident angle $\theta_{\rm inc}$:
\begin{equation}
\rm tan(\theta_t) \;=\;  \frac{\rm d \rm k_n}{\rm d \rm k_\tau} |_{\! \; \rm k_\tau=\sqrt{\epsilon_0}\frac{\omega}{c} \rm sin(\theta_{\; \! \rm inc})}.
\label{refraction_angle}
\end{equation}
To evaluate the dependance of $\theta_{\rm t}$ with respect to $\theta_{\rm inc}$, a relative refractive index $n^*$ is defined referring to Snell's law as
\begin{equation}
n^* \; = \;  \frac{\rm sin(\theta_t) }{\rm sin(\theta_{\rm inc}) }.
\label{index}
\end{equation}
With optimal layer thicknesses $d_1$ and $d_2$ satisfying Eq. (1-2), one finds $n^* = -1$ and Veselago lensing can be achieved.

As can be seen in Fig. 2(d), the value of $n^*$ of PHC with optimal $d_1$ and $d_2$ is almost constant for all incident angles $\theta_{\rm inc}$, while for hyperbolic medium (e.g. sapphire) it strongly depends on $\theta_{\rm inc}$. Thus a nearly-perfect focusing can be obtained by HPC even when material losses are included. For larger incident angles (corresponding to larger tangential momentum $k_\tau$), the refracted rays is partially deflected from the focus. Nevertheless, most energy of incoming waves is transferred by this Veselago lens and recovered near the predicted exit focus. 

%
%------------------------  Fig. 3 ---------------------
\begin{figure}[h]
\centering
\includegraphics[width=8.5cm]{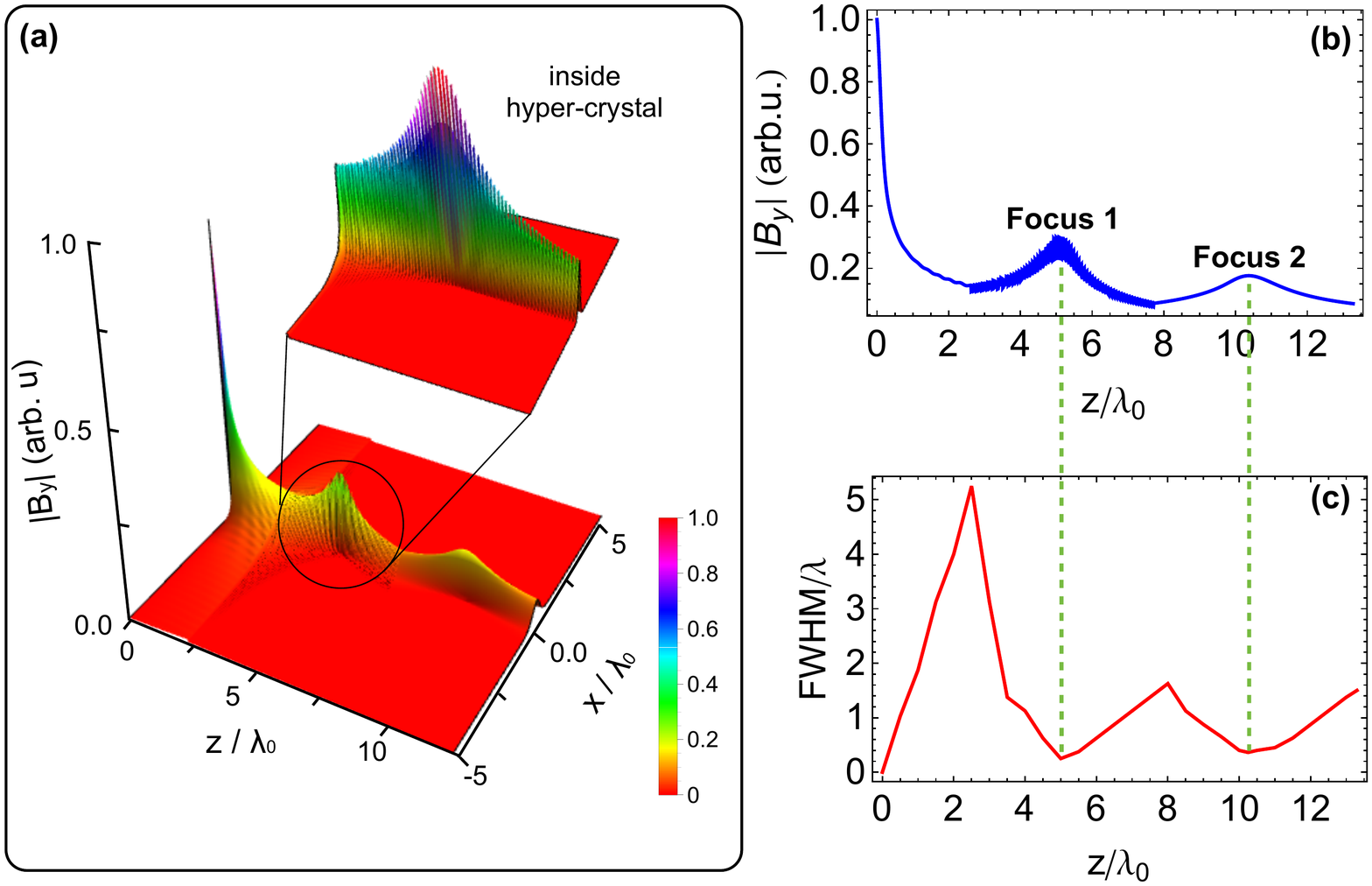} 
\noindent
\caption{\noindent (a) 3D profiles of $\rm |B_y|$ of a TM polarized slot source. The inset shows enlarged $\rm |B_y|$ distributions inside the proposed Veselago lens. (b) The peak magnitude of $\rm |B_y|$ at $x = 0$. A ``double-focus" effect is clearly shown signifying the Veselago lensing. (c) The variation of transverse FWHMs of $\rm |B_y|$ in units of wavelength in Si ($\lambda$) along the z-axis. The maximum FWHM occurs at the entrance plane of the Veselago lens, and two minimum FWHMs ($\sim \lambda/2$) can be observed near the two foci. Widening of the blue line in panel (b) corresponds to rapid oscillations of magnetic field at the range $2.5 < \rm z/\lambda_0 < 7.5$ due to the microstructure in PHC. 
}
\label{fig3}
\end{figure}
%------------------------------------------------------
%

To further confirm the typical ``double-focus" effect of our Veselago lens (two foci formed inside an outside the lens) and assess the quality of imaging, we show the 3D distribution of electromagnetic field from a TM polarized point slot source at a distance around $2.5 \lambda_0$ from the lens. From Fig. 3, the double-focus effect is clearly observed and the peak positions of magnetic field $\rm |B_y|$ agree well with the theoretical prediction. Our Veselago lens also demonstrates a substantially improved image quality than those realized by PCs and HMMs. As shown in Fig. 3(c), the full width at half maximum (FWHM) of $\rm |B_y|$ in transverse direction (x-axis) at the outside focus is approximately 0.5$\lambda$, where $\lambda$ is the wavelength in Si ($\lambda=\lambda_0/3.42$). Although the peak $\rm |B_y|$ at the external focus shows a much wider FWHM along z-axis due to the material absorption in HPC, it might be favorable for detectors to locate the position of focus.

%%******************************************************  Sec. III  ****************************************************** %%
%\section{Discussion}
Note that the proposed Veselago lens does not amplify the evanescent waves, therefore, its resolution is defined by the conventional diffraction limit in the material that surrounds the device. In our example, the resolution corresponding to the minimum FWHM of $\rm |B_y|$ at the focus is around 0.5$\lambda$.

Our Veselago lens is based on the essential concept of PHC by controlling the dispersion with phase space bandgaps. This approach is fundamentally different from the recent work by Xu et. al. \cite{XuNature2013}, where Veselago lensing at ultraviolet is achieved by a bulk stack of strongly coupled plasmonic waveguides with an optimized unit cell of ``metal-dielectric-metal-dieletric-metal". The length of unit cell is on the order of half wavelength and an in-plane mode symmetry is required to obtain the omnidirectional left-handed response. 

Compared with Veselago lens realized by PCs or HMMs, our system shows significantly improved focusing and imaging quality. Due to the substantially reduced loss by Bragg scattering in PHC, the transmission efficiency is also enhanced. However, the phase refraction is still positive ($\vec{\bold S} \cdot \vec{\bold k} > 0$)  and our system doesn't exhibit a 3D isotropic negative refractive index. Moreover, since the single period of PHC is well below the free-space wavelength, a large number of alternating layers is needed for far-field imaging, which brings extra difficulties in fabrication. 

The value of $n^*$ of our proposed Veselago lens is close to -1, which matches the ambient Si. Therefore, this imaging system can be used for the detection of hot-spots in Si-based integrated circuits, especially when the hot-spots are located deeply inside a multi-layered semiconductor circuit or device. This method may greatly facilitate the failure analysis of semiconductor devices caused by local heating.  

%%******************************************************  Sec. IV  ****************************************************** %%
%\section{Conclusions}

In conclusion, we have shown that Veselago lens can be realized by planar PHC with a nearly-constant relative index of refraction. This Veselago lens can be implemented by a stacked structure of interleaved multilayers of dielectric and hyperbolic materials, which shows a nearly-perfect focus with much reduced image aberrations. It can find many engineering applications, such as the improvement of the photolithography and hot-spot detection in Si-based integrated circuits.  

This work was partially supported by Gordon and Betty Moore Foundation, NSF Center for Photonic and Multiscale Nanomaterials, and ARO MURI.

%%******************************************************  References ****************************************************** %%

\end{document}